\documentclass[aps,prb,twocolumn,superscriptaddress,showpacs]{revtex4-1}

\usepackage{graphicx}
\usepackage{calc}
\usepackage{bm}
\usepackage{color}
\usepackage{textcomp}

\begin{document}

\title{Surface superconductivity in a three-dimensional Cd$_3$As$_2$ semimetal at the interface with gold contact}

\author{O.O.~Shvetsov}
\affiliation{Institute of Solid State Physics of the Russian Academy of Sciences, Chernogolovka, Moscow District, 2 Academician Ossipyan str., 142432 Russia}
\affiliation{Moscow Institute of Physics and Technology, Institutsky lane 9, Dolgoprudny, Moscow region, 141700 Russia}
\author{V.D.~Esin}
\author{A.V.~Timonina}
\author{N.N.~Kolesnikov}
\author{E.V.~Deviatov}
\affiliation{Institute of Solid State Physics of the Russian Academy of Sciences, Chernogolovka, Moscow District, 2 Academician Ossipyan str., 142432 Russia}

\date{\today}

\begin{abstract}
    We experimentally investigate charge transport through a single planar junction between Cd$_3$As$_2$ Dirac semimetal and a normal Au lead. For non-superconducting bulk Cd$_3$As$_2$ samples,  we observe non-Ohmic $dV/dI(V)$ curves, which strongly resemble standard Andreev reflection with well-defined superconducting gap.  Andreev-like behavior is demonstrated for Cd$_3$As$_2$ samples with different surface and contact preparation techniques.   We connect this  behavior with surface superconductivity  due to the flat-band formation in Cd$_3$As$_2$, which has been predicted theoretically. The conclusion on superconductivity is also supported by the gap suppression by magnetic fields or temperature.
\end{abstract}

\pacs{73.40.Qv  71.30.+h}

\maketitle

\section{Introduction}

    Cd$_3$As$_2$ is predicted to be a three-dimensional Dirac semimetal~\cite{cdasreview}, so it has symmetry-protected conic dispersion in the bulk spectrum~\cite{wang1, wang2}, which has been experimentally confirmed by angle-resolved photoemission spectroscopy (ARPES)~\cite{arpes1, arpes2} and scanning tunneling microscopy~\cite{stm} measurements. Due to the Dirac spectrum, Cd$_3$As$_2$ demonstrates interesting physical properties, e.g. unusual magnetoresistance phenomena, associated with chiral anomaly~\cite{phe1, phe2}, and ultrahigh carrier mobility~\cite{mobility1, mobility2}. Some features of exotic surface transport have been demonstrated by observation of quantum oscillations~\cite{osc}.

    By breaking certain symmetries,  Cd$_3$As$_2$ can be driven to different topological phases~\cite{wang1}, such as topological insulator~\cite{ti}, Weyl semimetal~\cite{weyl1,weyl2}, or even topological superconductor~\cite{tsc1,tsc2,tsc4}. The latter is notably attractive due to the surface states hosting Majorana fermions~\cite{tsc2,mf1,mf2,mf3}.

    There are two ways to induce superconductivity in bulk Cd$_3$As$_2$: by carrier doping~\cite{wang1}, which is only a theoretical prediction so far, or by external pressure~\cite{pressure}. In the latter case bulk superconductivity appears~\cite{pressure} around 3.5 GPa. In addition, point contact spectroscopy experiments~\cite{tpc1,tpc2} reveal signatures of superconductivity in a tip contact region (so-called tip induced superconductivity), while no effect is observed in the case of a soft contact~\cite{tpc2}. The origin of the effect is still debatable, e.g., it is not clear, whether pressure of a tip is enough to induce superconductivity in Cd$_3$As$_2$.

On the other hand, flat-band formation  stimulates surface superconductivity~\cite{barash,kopnin2,flatTc1,flatTc2}. In the presence of attractive interaction due to electron-phonon coupling~\cite{pressure}, the high density of states associated with these flat bands dramatically increases the superconducting transition temperature. This property is generic and does not depend much on the details of the system~\cite{kopnin2}. In particular, superconductivity has been observed in twisted bilayer graphene~\cite{graphene1,graphene2,graphene3}. 

Flat bands may emerge due to  interaction~\cite{khodel1,khodel2} or  topological effects~\cite{kopnin2,volovik1,kopnin1,mikitik}.  Historically flat bands were first discussed in the context of Landau levels. Now, they are considered  as  a class of fermionic systems with a dispersionless spectrum that has exactly zero energy, i.e. with diverging density of states. Interaction effects could be expected for high-mobility carriers~\cite{cdasreview} in  Cd$_3$As$_2$. The simplest example of topological flat-band formation is known for nodal-line semimetals~\cite{volovik1,kopnin1}. On the boundary of each topological insulator inside the nodal loop there should be the zero energy state. But this occurs for all the insulators inside the loop, so all these zero energy states on the surface form the 2D flat band. The topological flat-band formation is not also impossible for Cd$_3$As$_2$ material, since it is known to experience transition to different topological phases~\cite{wang1,silicon,strain}.  It is important that  if surface superconductivity appears in Cd$_3$As$_2$ Dirac semimetal due to fundamental effects, it should be independent on the contact preparation technique.  

Here, we experimentally investigate charge transport through a single planar junction between Cd$_3$As$_2$ Dirac semimetal and a normal Au lead. For non-superconducting bulk Cd$_3$As$_2$ samples  we observe non-Ohmic $dV/dI(V)$ curves, which strongly resemble standard Andreev reflection with well-defined superconducting gap.  Andreev-like behavior is demonstrated for Cd$_3$As$_2$ samples with different surface and contact preparation techniques.   We connect this  behavior with surface superconductivity  due to the flat-band formation in Cd$_3$As$_2$, which has been predicted theoretically. The conclusion on superconductivity is also supported by the gap suppression by magnetic field or temperature.

\section{Samples and technique}

\begin{figure}
\centerline{\includegraphics[width=\columnwidth]{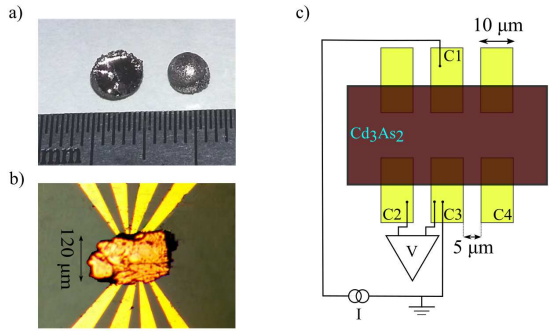}}
\caption{(Color online) (a) An initial Cd$_3$As$_2$ drop (right) and a one cleaved along (112) (left). (b) Top-view image of the sample with a small Cd$_3$As$_2$ single crystal (c) The sketch of a sample with electrical connections. 100 nm thick and 10 $\mu$m wide Au leads are formed on a SiO$_2$ substrate. A Cd$_3$As$_2$ single crystal ($\approx$ 100~$\mu$m size)  is transferred on top of the leads with $\approx$ 10 $\mu$m overlap, forming planar junctions. Charge transport is investigated with a standard three-point technique: the studied contact (C3) is grounded and two other contacts (C1 and C2) are used for applying current and measuring potential. }
\label{cdas_sample}
\end{figure}

Cd$_3$As$_2$ crystals were grown by crystallization of molten drops in the convective counterflow of argon held at 5 MPa pressure. For the source of drops the stalagmometer similar to one described~\cite{growth} was applied. The crystals  sometimes had signs of partial habit of $\alpha$-Cd$_3$As$_2$ tetragonal structure. About one fifth of the drops were  single crystals, like ones depicted in Fig.~\ref{cdas_sample} (a). The EDX measurements and X-ray powder diffractograms always confirmed pure Cd$_3$As$_2$.

Fig.~\ref{cdas_sample} (b) shows a top-view image of a sample. The leads pattern is formed by lift-off technique after thermal evaporation of 100 nm Au on the insulating SiO$_2$ substrate. The 10~$\mu$m wide Au leads are  separated by 5~$\mu$m intervals, see  Fig.~\ref{cdas_sample} (b). 

 Small (less than 100 $\mu$m size) Cd$_3$As$_2$ single crystals  are obtained by a mechanical cleaving method, somewhat similar to described in Ref.~\onlinecite{yu}: we crush the initial 5~mm size Cd$_3$As$_2$ single crystal onto small fragments. This procedure allows to create a clean Cd$_3$As$_2$ surface  without mechanical polishing or chemical treatment. 

Then, the obtained small Cd$_3$As$_2$  crystal is transferred to the Au leads pattern and pressed slightly with another oxidized silicon substrate.  A special metallic frame allows to keep substrates parallel and apply a weak pressure to the piece. No external pressure is needed for a Cd$_3$As$_2$ crystal to hold on a substrate with Au leads afterward.  

For comparison, we also defined 100 $\mu$m $\times$ 100 $\mu$m  Au contacts by standard photolithography on the cleaved  along (112) and mechanically polished  surface of the initial   Cd$_3$As$_2$ drop. In this case, Cd$_3$As$_2$ surface degradation could be expected due to the  polishing process~\cite{crassee}.

\begin{figure}
\centerline{\includegraphics[width=0.8\columnwidth]{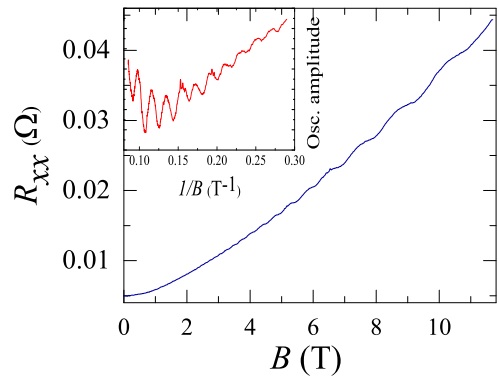}}
\caption{(Color online)  Transverse four-point magnetoresistance with Shubnikov de Haas oscillations in high magnetic fields~\protect\cite{cdasreview} for one of our samples at 60 mK. The ac measurement current is 2.5~$\mu$A at 110~Hz, the ac $xx$ voltage is measured by lock-in after a preamplifier. Inset demonstrates perfect periodicity of the oscillations in the inverse magnetic field. The data in the inset are obtained by subtracting a linear dependence from the raw $R(B)$ curve, shown in the main figure.}
\label{magres}
\end{figure}

We check by standard magnetoresistance measurements that our Cd$_3$As$_2$ samples demonstrate large magnetoresistance with Shubnikov de Haas oscillations in high magnetic fields~\cite{cdasreview}, see  Fig.~\ref{magres}, indicating high quality of Cd$_3$As$_2$. From the oscillations' period in the inverse magnetic field, see the inset, and zero-field resistance value we estimate the concentration of carries as  $n \approx $ 2.3$\times$10$^{18}$~cm$^{-3}$ and low-temperature mobility as $\mu \approx$ 10$^6$~cm$^2$/Vs, which is in the good correspondence with known values~\cite{cdasreview}.

We study electron transport across a single Au-Cd$_3$As$_2$ junction in a standard three-point technique, see Fig.~\ref{cdas_sample} (a): one Au contact is grounded and two other contacts are used for applying current and measuring Cd$_3$As$_2$ potential. To obtain $dV/dI(V)$ characteristics, dc current is additionally modulated by a low (below the dc points step) ac component. We measure both dc ($V$) and ac ($\sim dV/dI$) components of the potential with a dc voltmeter and a lock-in, respectively. We check, that the lock-in signal is independent of the modulation frequency. The measurements are performed in a dilution refrigerator equipped with superconducting solenoid.

\section{Experimental results}

\begin{figure}
\includegraphics[width=\columnwidth]{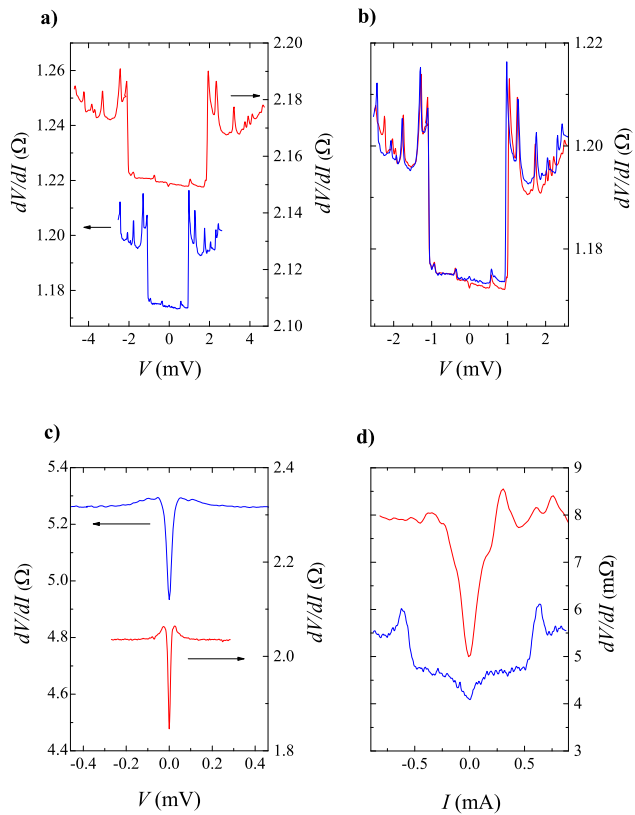}
\caption{(Color online) Examples of $dV/dI(V)$ characteristics are shown in Fig.~\ref{IV} for different Au-Cd$_3$As$_2$ junctions. The main experimental finding is the prominent non-Ohmic behavior, which is reflected as about 10\% resistance dip around zero bias and is well reproducible for different samples. (a) $dV/dI(V)$ characteristics  for the same Au-Cd$_3$As$_2$ junction, obtained for  two different voltage probes, which allows to estimate the bulk resistance contribution. (b) Coincidence of the curves from (a) after subtracting the bulk resistance (see the text).  Comparison of (a) and  (c) demonstrates maximum device-to-device fluctuations for junctions with cleaved  Cd$_3$As$_2$ fragments, while the panel (d) shows $dV/dI(V)$ curve variation for  polished Au-Cd$_3$As$_2$ junctions.    The curves are obtained at 60~mK in zero magnetic field.}
\label{IV}
\end{figure}

Examples of $dV/dI(V)$ characteristics are shown in Fig.~\ref{IV} for different Au-Cd$_3$As$_2$ junctions. Since the $dV/dI(V)$ curves of the junctions might be sensitive to the interface quality, e.g. as it is known for normal-superconductor junctions~\cite{BTK},  Fig.~\ref{IV} (a) and (c) demonstrate maximum device-to-device fluctuations for samples with cleaved  Cd$_3$As$_2$ fragments, while Fig.~\ref{IV} (d) shows $dV/dI(V)$ curve variation for large junctions on polished Cd$_3$As$_2$ surface. 

The main experimental finding is the prominent non-Ohmic behavior, which is reflected in about 10\% resistance dip around zero bias. This  behavior is well-reproducible for different samples, see Fig.~\ref{IV}: while the shape and the width of the dip may vary from sample to sample, the qualitative behavior is the same.

$dV/dI(V)$ characteristics are shown in Fig.~\ref{IV} (a)  for two different positions of a voltage probe for the same Au-Cd$_3$As$_2$ junction. In the three-point configuration, the measured potential reflects the in-series connected resistances of the grounded Au lead, the Au-Cd$_3$As$_2$ interface, and the bulk resistance of Cd$_3$As$_2$. If one changes only the voltage probe position, two former contributions are invariant. Only the contribution of the bulk Cd$_3$As$_2$ resistance is varied, which we detect as the resistance  level change $\delta R$ in Fig.~\ref{IV} (a). The curves coincide with high accuracy after subtracting $\delta R$ and $I \delta R$ from $dV/dI$ and $V$ components of the upper curve,  see  Fig.~\ref{IV} (b), so the $dV/dI$ dip does not originate from the Cd$_3$As$_2$ bulk. We should relate the resistance dip with Au-Cd$_3$As$_2$ interface contribution, since  non-Ohmic $dV/dI(V)$ behavior can not be linked with   Au lead.

It is clear, that $dV/dI(V)$ characteristics of Au-Cd$_3$As$_2$ interface strongly resemble standard Andreev reflection behavior~\cite{andreev,tinkham} for transparent normal-superconducting junctions. This conclusion is supported by $dV/dI(V)$ nonlinearity suppression by magnetic field or temperature: despite the different shape of the original $dV/dI(V)$ curves  in  Fig.~\ref{IV}, all of them become flat above some critical temperature or magnetic field. 

We give an example of temperature and magnetic field evolution in Figs.~\ref{IV_T} and ~\ref{IV_B} for the junction from Fig.~\ref{IV} (a). 
The width of the $dV/dI$ dip is gradually diminishing, as it is shown in Fig.~\ref{IV_T} (b) and in the inset to Fig.~\ref{IV_B} as function of temperature and magnetic field, respectively. The behavior strongly resembles the known one~\cite{tinkham} for a superconducting gap,  but the data can not be fitted by standard BCS temperature dependence, and $(1-H^2/H^2_c)$ magnetic field law (the solid lines in Fig.~\ref{IV_T} (b) and in the inset to Fig.~\ref{IV_B}), known for the conventional superconductors~\cite{tinkham}. Thus, the  unconventional  superconductivity is possible, like it was proposed in  point contact spectroscopy experiments~\cite{tpc1,tpc2}.  The critical temperature  can be estimated as  $T_c\approx 1$~K in the inset to Fig.~\ref{IV_T} (note, that the curves in Fig.~\ref{IV} (a-b) still contain an unknown bulk contribution). For the samples in Fig.~\ref{IV} (c) and (d), we estimate $T_c$ as 300~mK and 1~K, $B_c$ as 26~mT and 140~mT, respectively. 

The bulk Cd$_3$As$_2$ material is not superconducting~\cite{pressure}, which is confirmed by finite four-point resistance in Fig.~\ref{magres}. Since an Au lead is also normal, the Andreev-like behavior of experimental $dV/dI(V)$ curves should reflects surface  superconductivity at Au-Cd$_3$As$_2$ interface.

\section{Discussion}

As a result, we observe $dV/dI(V)$ curves, which are qualitatively analogous to  tip induced superconductivity~\cite{tpc1,tpc2}, for wide planar contacts without external pressure. 

\begin{figure}
\centerline{\includegraphics[width=\columnwidth]{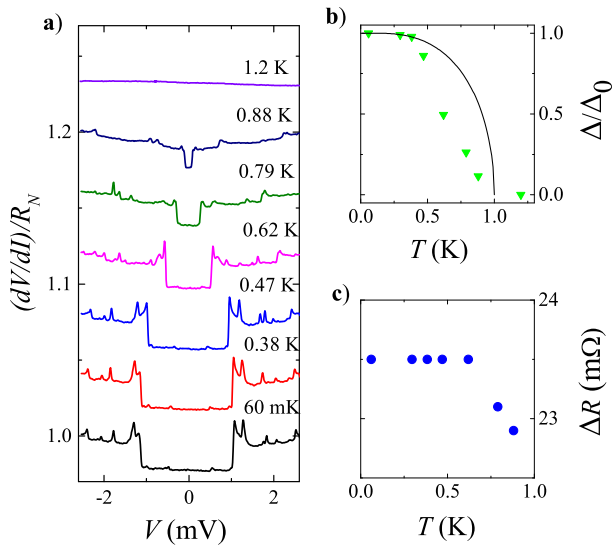}}
\caption{(Color online) (a) Suppression of $dV/dI(V )$ non-linearity by temperature.  The curves are shifted for clarity. 
(b) The gap $\Delta$, obtained as the non-linearity width at half of its' maximum depth, as function of temperature. The data resemble superconducting gap behavior, but they can not be fitted by standard BCS dependence~\cite{tinkham} (black line). (c) The depth $\Delta R$ of the resistance dip at zero bias, which is consistent with the BTK dependence~\cite{BTK,tinkham} for the transparent interface. The data are shown for zero magnetic field.}
\label{IV_T}
\end{figure}

Formally, standard BTK-theory~\cite{BTK} is appropriate in the ballistic limit, when the contact diameter is less than elastic and inelastic mean free paths. In the opposite (thermal) limit, peaks in $dV/dI$ reflect superconducting transition due to the critical current in the junction. 

\begin{figure}
\centerline{\includegraphics[width=0.65\columnwidth]{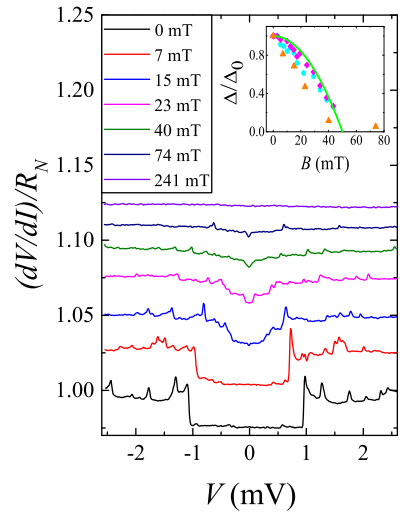}}
\caption{(Color online) Suppression of $dV/dI(V )$ non-linearity by normal magnetic field at 60~mK temperature.  The curves are shifted for clarity.
Inset demonstrates the gap   $\Delta$ (triangles), obtained as the non-linearity width at half of its' maximum depth, diminishing with magnetic field, as also expected for superconducting gap. The data can not be well fitted by standard dependence~\cite{tinkham} (green line), which may hint on the unconventional superconductivity~\cite{tpc1,tpc2}. The gap weakly depends on the field orientation, as it is shown for another sample  by  circles (in-plane field orientation) and diamonds (normal one).}
\label{IV_B}
\end{figure}

The ballistic regime is obviously realized for the clean Cd$_3$As$_2$ surface, as we see in Fig.~\ref{IV} (a) and (b), since the mean free path exceeds 25~$\mu$m at given concentration and mobility. 
In this case, in contrast to the tip experiments,  the width of the dip is defined by the superconducting gap for the best junctions, like in Fig.~\ref{IV} (a), as it is expected~\cite{BTK,andreev,tinkham} for standard Andreev reflection~\cite{ingasb,nbsemi}. This conclusion is supported by qualitative behavior of $\Delta(T)$ and $\Delta(B)$ dependencies in Fig.~\ref{IV_T} (b) and in the inset to Fig.~\ref{IV_B}. Also, the depth $\Delta R$ of the resistance dip is nearly constant at low temperatures, see Fig.~\ref{IV_T} (c),  which is consistent with the BTK dependence~\cite{BTK,tinkham} for the transparent interface. We wish to emphasize here, that the actual gap value is smaller than the width of the dip in Figs.~\ref{IV_T},~\ref{IV_B} because the $dV/dI(V)$ curves in Fig.~\ref{IV} (a-b) still contain an unknown bulk contribution. The actual gap value should be obtained from the $\Delta(T)$ dependence.

The thermal limit is obviously realized for the polished Cd$_3$As$_2$ surface with large contacts in Fig.~\ref{IV} (d), so the differential resistance is driven by current, which achieves the critical value $I_c$ at low (about 1~$\mu$V) imbalances at the interface. We wish to mention, that for narrow superconductors between two massive normal metals, electron cotunneling and crossed Andreev reflection should be taken into account~\cite{car}. Both these effects are extremely sensitive to the transmission of the interfaces, see Fig.~12 in Ref.~\onlinecite{car}, which should be responsible for the device-to-device fluctuations in  Fig.~\ref{IV} (a) and (c).

The interface superconductivity has been demonstrated in a number of various systems, with different discussed microscopic mechanisms~\cite{surf_review}. In our experiment,  superconductivity should originate from fundamental  effects  in topological Dirac semimetal, since it is independent of  contact preparation details. The obvious candidate is the flat-band formation~\cite{barash,kopnin2,flatTc1,flatTc2} due to interaction or topology. For our samples with the bulk carrier density $n \approx $ 2.3$\times$10$^{18}$~cm$^{-3}$ and the corresponding~\cite{cdasreview} effective mass 0.044, the interaction parameter $r_s$ is about 1. Even if this value is  enhanced for low densities near the sample surface, it seems to be too small to produce noticeable interaction effects~\cite{khodel1,khodel2}.  On the other hand, Cd$_3$As$_2$ Dirac semimetal  is known to experience transition to different topological phases~\cite{wang1}, so one could propose topological mechanism, similar to surface states in nodal-line semimetals~\cite{kopnin2,volovik1,kopnin1,mikitik}. The possibility of such transitions is also supported by recent theoretical predictions for different semimetal systems~\cite{silicon}. For Cd$_3$As$_2$ Dirac semimetal,  flat bands are evidenced in ARPES~\cite{neu,roth}  and magneto-optics~\cite{hakl,akrap} experiments.

Another possibility is the strain effects. In Dirac semimetals strain generically acts as an effective gauge field on Dirac fermions and creates pseudo-Landau orbitals without breaking time-reversal symmetry~\cite{strain}. The zero-energy Landau orbitals form a flat band in the vicinity of the Dirac point, so the high density of states of this flat band gives rise to interface superconductivity. We observe finite four-point resistance between different contacts in Fig.~\ref{magres}, which well correspond to the fact, that strain-induced flat-band formation is only occurs   at  Au-Cd$_3$As$_2$ interface due to materials misfit.

\section{Conclusion}
As a conclusion, we experimentally investigate charge transport through a single planar junction between Cd$_3$As$_2$ Dirac semimetal and a normal Au lead. For non-superconducting bulk Cd$_3$As$_2$ samples,  we observe non-Ohmic $dV/dI(V)$ curves, which strongly resemble standard Andreev reflection with well-defined superconducting gap.  Andreev-like behavior is demonstrated for Cd$_3$As$_2$ samples with different surface and contact preparation techniques.   We connect this  behavior with surface superconductivity  due to the flat-band formation in Cd$_3$As$_2$, which has been predicted theoretically. The conclusion on superconductivity is also supported by the gap suppression by magnetic fields or temperature.

\acknowledgments
We wish to thank A.~Kononov for help with experimental setup, G.E.~Volovik, Yu.S. Barash, and V.T.~Dolgopolov for fruitful discussions.  We gratefully acknowledge financial support partially by the RFBR  (project No.~19-02-00203), RAS, and RF State task.

\end{document}